\documentclass[10pt,a4paper,english]{article}\usepackage[]{graphicx}\usepackage[]{color}
\makeatletter
\def\maxwidth{ %
  \ifdim\Gin@nat@width>\linewidth
    \linewidth
  \else
    \Gin@nat@width
  \fi
}
\makeatother

\definecolor{fgcolor}{rgb}{0.345, 0.345, 0.345}

\usepackage{framed}
\makeatletter
 {\par\unskip\endMakeFramed%
 \at@end@of@kframe}
\makeatother

\definecolor{shadecolor}{rgb}{.97, .97, .97}
\definecolor{messagecolor}{rgb}{0, 0, 0}
\definecolor{warningcolor}{rgb}{1, 0, 1}
\definecolor{errorcolor}{rgb}{1, 0, 0}

\usepackage{alltt}

\usepackage[hyphens]{url}
\usepackage{amsmath}
\usepackage{amsfonts}
\usepackage{amssymb}
\usepackage{amsthm}

\theoremstyle{plain}

\theoremstyle{definition}

\theoremstyle{remark}

\PassOptionsToPackage{hyphens}{url}\usepackage{hyperref}
\usepackage{hyperref}
\usepackage[square,numbers]{natbib}

\usepackage[newitem,newenum,increaseonly]{paralist}


\usepackage[notheorems]{common}

\providecommand{\util}[1]{\funcitUL{U}{}{}{#1}}
\providecommand{\dutil}[2][{}]{\funcitUL{U}{\wrapNeParens{#1}}{}{#2}}

    \usepackage{color} 
    \usepackage{enumerate} 
    \usepackage{amsmath} 
    \usepackage{amssymb} 
		\usepackage{fancyvrb} 
    \usepackage{hyperref}

    \definecolor{orange}{cmyk}{0,0.4,0.8,0.2}
    \definecolor{darkorange}{rgb}{.71,0.21,0.01}
    \definecolor{darkgreen}{rgb}{.12,.54,.11}
    \definecolor{myteal}{rgb}{.26, .44, .56}
    \definecolor{gray}{gray}{0.45}
    \definecolor{lightgray}{gray}{.95}
    \definecolor{mediumgray}{gray}{.8}
    \definecolor{inputbackground}{rgb}{.95, .95, .85}
    \definecolor{outputbackground}{rgb}{.95, .95, .95}
    \definecolor{traceback}{rgb}{1, .95, .95}
    \definecolor{red}{rgb}{.6,0,0}
    \definecolor{green}{rgb}{0,.65,0}
    \definecolor{brown}{rgb}{0.6,0.6,0}
    \definecolor{blue}{rgb}{0,.145,.698}
    \definecolor{purple}{rgb}{.698,.145,.698}
    \definecolor{cyan}{rgb}{0,.698,.698}
    \definecolor{lightgray}{gray}{0.5}
    
    \definecolor{darkgray}{gray}{0.25}
    \definecolor{lightred}{rgb}{1.0,0.39,0.28}
    \definecolor{lightgreen}{rgb}{0.48,0.99,0.0}
    \definecolor{lightblue}{rgb}{0.53,0.81,0.92}
    \definecolor{lightpurple}{rgb}{0.87,0.63,0.87}
    \definecolor{lightcyan}{rgb}{0.5,1.0,0.83}

    \definecolor{incolor}{rgb}{0.0, 0.0, 0.5}
    \definecolor{outcolor}{rgb}{0.545, 0.0, 0.0}

    \hypersetup{
      breaklinks=true,  
      colorlinks=true,
      urlcolor=blue,
      linkcolor=darkorange,
      citecolor=darkgreen,
      }
    
\IfFileExists{upquote.sty}{\usepackage{upquote}}{}
\begin{document}

\title{Safety Third: Roy's Criterion and Higher Order Moments}
\author{Steven E. Pav \thanks{\email{spav@alumni.cmu.edu}}}

\maketitle

\begin{abstract}
Roy's `Safety First' criterion for selecting one risky asset from
many is adapted to the case of non-normal returns, via Cornish Fisher expansion. 
The resulting investment objective is consistent with first order stochastic 
dominance, and is equal to the \txtSR for the case of normal returns. 
An investor selecting assets via this objective is not universally attracted 
to positive skew, rather the preference for skew depends on term, 
the expected return and the disastrous rate of return.
\end{abstract}

\section{Introduction}

Mathematical economic theory posits that agents seek to maximize some
utility function.  \cite{eeckhoudt2005economic} 
In practice, however, real investors can rarely evoke 
their own utility functions. Rather, when selecting from a number of risky
assets, investors (and quantitative-minded asset managers) often rank their
choices based on the moments of the returns stream, preferring \eg 
higher expected returns for a fixed level of volatility, \cetpar.
Arguably the most commonly used measure of investment opportunities is the
\emph{\txtSR}, here defined as $\psnr = \fracc{\wrapParens{\pmu - \rfr}}{\psig},$ 
where \rfr is the `disastrous' or `risk-free' rate of return, and 
\pmu and \psigsq are the expected value and variance of the returns stream,
assumed to be known\footnote{It might be more accurate to call \psnr the \txtSNR, 
and reserve the term \txtSR for the analogous quantity constructed from sample
estimates. Sharpe himself notes, ``Since the predictions cannot be obtained in
any satisfactory manner, \ldots ex post values must be used--the average rate
of return of a portfolio must be substituted for its expected rate of return,
and the actual standard deviation of its rate of return for its predicted 
risk.'' \cite[p. 122]{Sharpe:1966} However, we will follow common usage in
calling \psnr the \txtSR, without much risk of confusion.}.

One objection to the use of the \txtSR as an investment objective is 
that it is generally not consistent with first order stochastic dominance. 
\cite{hodges1998generalization,NBERw19500,zakamulin2008portfolio}
That is, one can construct two random variables, say \reti[] and \retj[],
such that \reti[] stochastically dominates \retj[], but the \txtSR
of \reti[] is lower than that of \retj[]. 
Moreover this deficiency cannot be solved by assuming away
the $\pmu < 0$ case\footnote{The \txtSR as an objective `prefers' higher volatility
in the case $\pmu < 0$, and is thus clearly inconsistent with second-order
stochastic dominance. It is not clear, however, that the sample analogue shares
this deficiency.}
Hodges' provides the classical counterexample, but such pathological cases are
easy to construct, as shown in the appendix. 

There have been numerous attempts to generalize the \txtSR to remedy these
deficiencies, making it suitable for the case of non-normal returns by including
higher order moments.  \cite{hodges1998generalization,NBERw19500,zakamulin2008portfolio}
Hodges assumes an investor with the CARA utility function, $\util{w} = -
\exp{-\lambda w}$. For an asset with normally distributed returns, the optimal
amount to invest, long or short, in the asset is\footnote{\nb this is
essentially the \txtMP on one asset.}
$\fracc{\pmu}{\lambda\psigsq}$, in the sense of maximizing the \emph{expected}
utility. The maximum expected utility at this allocation is 
$-\exp{-\half \wrapParens{\frac{\pmu}{\psig}}^2}$, ignoring the time term for
simplicity. This leads Hodges to define the ``Generalized \txtSR'' as 
\begin{equation}
\psnrg = \sqrt{-2 \flog{-U^{*}}}.
\end{equation}
where $U^{*}$ is the maximum expected utility under the CARA utility function.
\cite{hodges1998generalization}
That is
\begin{equation}
U^{*} \defeq \max_x \E{- \exp{-\lambda x w}},
\end{equation}
and so
\begin{equation}
\psnrg = \sqrt{\max_x -2\flog{\E{\exp{-\lambda x w}}}}.
\end{equation}

As Hodges' objective is difficult to compute, Zakamouline and Koekebakker carry
his analysis to its logical conclusion, using Taylor's theorem to describe
the Generalized \txtSR in terms of investor's relative preferences for higher
order moments of wealth.  \cite{zakamulin2008portfolio} They derive an 
``adjusted for skew \txtSR'', defined as
\begin{equation}
\psnr[3] = \psnr\sqrt{1 + b_3 \frac{\pzkuml[3]}{3} \psnr},
\end{equation}
where \pzkuml[3] is the skewness of the returns distribution, and $b_3$ is the
investor's relative preference for third order moments:
\begin{equation*}
b_3 = \frac{a_3}{a_2^2},\quad\mbox{where}\quad a_k = \frac{\dutil[k]{w_r}}{\dutil[1]{w_r}},
\end{equation*}
and \dutil[k]{w_r} denotes the \kth{k} derivative of the investor's utility
function at the zero dollar allocation in the risk asset, denoted as $w_r$. For
an investor with HARA utility, the quantity $b_3$ is generally positive, and
thus the skew adjusted \txtSR has positive derivative with respect to skewness
(assuming $\psnr > 0$). In fact, a necessary condition for the investor to
demonstrate decreasing risk aversion is that $b_3 \ge 1$, a result due to
Pratt. \cite{zakamulin2008portfolio,pratt1964}

Smetters and Zhang carry this line of analysis further, showing that 
a valid ranking of investments must take into account investor's preferences
and cannot be a function only of the distributions of returns.  \cite{NBERw19500}
Moreover, they develop a ranking measure like the \txtSR expressed in terms of the cumulants
of the returns distribution and the derivatives of the utility.  
Their Theorem 9 establishes positive derivative of their objective with respect to odd
cumulants and negative derivative with respect to even cumulants of the returns
distribution, in accordance with the usual interpretations of `temperance',
`prudence', `edginess', \etc \cite{NBERw19500,denuitprudence} 
Smetters and Zhang describe how to approximately compute their objective,
showing that their third order approximation matches that of Zakamouline and
Koekebakker.


It is only by Stigler's Law of Eponymy that we know the quantity \psnr as 
``the \txtSR,'' instead of ``Roy's criterion.''  \cite{stiglersLaw} 
Sharpe first described his ``reward-to-variability ratio'' in 1966 as a
yardstick for comparing mutual funds, but Roy described the same quantity in 
1952 as a means of choosing among risky assets, under the moniker of
``Safety First.''  \cite{Sharpe:1966,roySafety1952,royTheMan} 
Roy's justification for this objective followed from Chebyshev's inequality, 
which states that 
\begin{equation}
\label{eqn:chebyshevs}
\Pr{\abs{\reti[] - \pmu} \ge \sqrt{k}\psig} \le \frac{1}{k}.
\end{equation}
For a given $\rfr < \pmu$, let $\sqrt{k} = \wrapParens{\pmu - \rfr}/\psig$.
Then since 
$\Pr{\reti[] - \pmu \le - \sqrt{k}\psig} \le \Pr{\abs{\reti[] - \pmu} \ge
\sqrt{k}\psig}$, we have
\begin{equation}
\label{eqn:roy_jackpot}
\Pr{\reti[] \le \rfr} =
\Pr{\reti[] - \pmu \le - \frac{\pmu - \rfr}{\psig}\psig} 
\le \wrapParens{\frac{\psig}{\pmu - \rfr}}^2 = \oneby{\psnrsq}.
\end{equation}
Thus to minimize the probability of a loss (relative to \rfr), one
should maximize \psnr.


\section{Safety First}

The crux of Roy's justification for the `Safety-First' objective,
which is just the \txtSNR, is that it bounds the probability of a loss,
defined as a return less than \rfr. The argument, based on 
Chebyshev's inequality, is only a rough upper bound. 
There are some situations, however, where the \txtSNR is exactly monotonic
in the probability of a loss. For example, if the returns are drawn
from a scale-location family, like the Gaussian family.
Note that the central limit theorem tells us that, conditional on finite
variance, the sample mean of some random variable converges to a normal
distribution, and thus for the case of log returns, 
since the mean return is just the total log return rescaled, the long term 
log return is approximately drawn from a scale-location family.

We can maintain the spirit of Roy's criterion by directly optimizing
the quantity he sought to maximize, \viz the probability of exceeding \rfr. 
To match the \txtSR in the case of Gaussian returns, we need only invert the 
normal CDF, resulting in the quantity:
\begin{equation}
\label{eqn:def_hosnr_I}
\psnrh\defeq -\pinorm[\Pr{\reti[] \le \rfr}],
\end{equation}
where \pnorm[\cdot] is the CDF of the normal distribution.
When $\reti[] \sim \normlaw{\pmu,\psigsq},$ the probability that $x \le \rfr$
is $\pnorm[\fracc{\wrapParens{\rfr - \pmu}}{\psig}]$, and so \psnrh equals the
\txtSR, \fracc{\wrapParens{\pmu - \rfr}}{\psig}. This objective is 
legitimately a `generalized \txtSR', since it agrees with the \txtSR exactly
for normal returns.  \cite{zakamulin2008portfolio}

It is trivial to verify that \psnrh is consistent with first order stochastic
dominance, or at least not inconsistent with it\footnote{This statement is 
weak, but cannot be strengthened; it must be admitted, for example, 
that for most \rfr, \psnrh makes no distinction between the two assets of 
Hodges' classic counterexample.}. 
Since if \reti[] stochastically dominates \retj[], 
$\Pr{\reti[] \le \rfr} \le \Pr{\retj[] \le \rfr}$ for all \rfr. By monotonicity
of \pinorm[\cdot], \psnrh is no smaller for \reti[] than \retj[]. It should be
clear, however, that the converse does not, indeed can not, hold: if \psnrh
is higher for \reti[] than \retj[], for a single \rfr, it need not be the case
that \reti[] stochastically dominates \retj[]. The simple proof is that since
stochastic dominance does not form a total ordering on probability distributions, 
but generalized Roy's criterion (for one choice of \rfr) does form a total ordering, 
the latter ordering cannot imply the former.

Roy's approximation is based on Chebyshev's inequality. We can construct
tighter approximations to the probability of a loss via
some classical approximations to the
central limit theorem. 
Suppose that one will observe \ssiz independent draws from the
returns stream, \reti. Without loss of generality\footnote{Here we assume the
returns are log returns. Then the sample mean is just the rescaled total
return. By similarly rescaling the disastrous return, we arrive at the
formulation here.}, let 
the disastrous event be that the observed sample mean return, \smu, is less than \rfr. 
This is equivalent to
$$
\sqrt{\ssiz}\frac{\smu - \pmu}{\psig} \le \sqrt{\ssiz}\frac{\rfr - \pmu}{\psig}.
$$
The cumulative distribution function of the quantity on the left hand
side can be approximated via some truncation of the
Edgeworth expansion.  \cite{edgeworthCF}

Define $\nctp \defeq \sqrt{\ssiz}\fracc{\wrapParens{\pmu - \rfr}}{\psig}$. 
The Edgeworth expansion is \cite[26.2.48]{abramowitz_stegun}
\begin{multline}
\label{eqn:edgeworth_exp}
\Pr{\sqrt{\ssiz}\frac{\smu - \pmu}{\psig} \le -\nctp} = 
\pnorm[-\nctp]
- \dnorm[\nctp] \wrapBracks{\frac{\pzkuml[3]}{6\sqrt{\ssiz}}\He[2]{\nctp}}\\
+ \dnorm[\nctp] \wrapBracks{\frac{\pzkuml[4]}{24\ssiz}\He[3]{\nctp}
 + \frac{\pzkuml[3]^2}{72\ssiz}\He[5]{\nctp}}\\
- \dnorm[\nctp] \wrapBracks{\frac{\pzkuml[5]}{120\ssiz^{\halff[3]}}\He[4]{\nctp}
 + \frac{\pzkuml[3]\pzkuml[4]}{144\ssiz^{\halff[3]}}\He[6]{\nctp}
 + \frac{\pzkuml[3]^3}{1296\ssiz^{\halff[3]}}\He[8]{\nctp}}\ldots
\end{multline}
where \pnorm and \dnorm are the cumulative distribution and density
functions of the standard unit normal, \He[i]{x} is the probabilist's
Hermite polynomial \cite[26.2.31]{abramowitz_stegun}, and 
\pzkuml[i] is the standardized \kth{i} cumulant, defined as the 
\kth{i} cumulant of the distribution divided by $\psig^i$. It happens
to be the case that \pzkuml[3] is the skewness, and \pzkuml[4] is the
excess kurtosis of the distribution.

Truncating beyond the $\ssiz^{-\halff}$ term and applying basic facts
of probability yields
\begin{equation}
\label{eqn:edgeworth_exp_II}
\begin{split}
\Pr{\smu \ge \rfr}&\approx
\pnorm[\nctp]
+ \frac{\dnorm[\nctp]}{\sqrt{\ssiz}}
	\wrapBracks{\frac{\pzkuml[3]}{6}\wrapParens{\nctp^2 - 1}}.\\
\end{split}
\end{equation}
The implication is that the probability that \smu exceeds \rfr will 
be increased if \nctp is large.  Moreover, for a fixed \nctp, the 
probability that \smu exceeds \rfr is increased
for large positive skew if $\nctp^2 > 1$, but for large negative
skew when when $\nctp^2 < 1$. The implication is that
when $\nctp^2$ is `large' (larger than one unit), one has positive
preference for skewed returns, otherwise one has negative preference.
As long as $\pmu > \rfr$, this is asymptotically compatible as $\ssiz\to\infty$
with the commonly held belief that investors universally value positive skew.

\subsection{Approximating Roy's criterion}

The generalized Roy's criterion of \eqnref{def_hosnr_I} is now expressed as
\begin{equation}
\label{eqn:def_hosnr_II}
\psnrh\defeq -\oneby{\sqrt{\ssiz}}
\pinorm[\Pr{\sqrt{\ssiz}\frac{\smu - \pmu}{\psig} \le -\nctp}].
\end{equation}
This implicit definition is a bit unwieldy for
use as an objective. One would prefer a definition in terms of the
cumulants of the returns stream. Rather than use the Taylor series expansion
of \pinorm, one can instead use the Cornish Fisher expansion of the sample 
quantile.  \cite{AS269,Jaschke01,CFetc}

Let $Y = \sqrt{\ssiz}\fracc{\wrapParens{\smu - \pmu}}{\psig}$. 
This is a random variable with zero mean and unit standard deviation. 
Let \pzkuml[i] be the \kth{i} standardized cumulant of 
\reti. The \kth{i} standardized cumulant of $Y$ is 
$\ssiz^{1-\halff[i]}\pzkuml[i]$.
The Cornish Fisher expansion \cite[26.2.49]{abramowitz_stegun}
finds $w$ in 
$$
\Pr{Y \le w} = \pnorm[z],
$$
in terms of $z$ and the higher order cumulants of the distribution. Setting
$w = -\nctp$, we have $z = -\sqrt{\ssiz}\psnrh$, and the Cornish Fisher
expansion reduces to 
\begin{equation}
\label{eqn:approx_hosnr}
\begin{split}
\psnrh &= \frac{\wrapParens{\pmu - \rfr}}{\psig} 
+ \oneby{{\ssiz}}\wrapBracks{\frac{\pzkuml[3]}{6}\He[2]{\sqrt{\ssiz}\psnrh}}\\
&\phantom{=}\,\,- \oneby{\ssiz^{\halff[3]}}\wrapBracks{\frac{\pzkuml[4]}{24}\He[3]{\sqrt{\ssiz}\psnrh}
-\frac{\pzkuml[3]^2}{36}\wrapBracks{2\He[3]{\sqrt{\ssiz}\psnrh}+ \He[1]{\sqrt{\ssiz}\psnrh}}}\\
&\phantom{=}\,\,+
\oneby{\ssiz^{2}}\left[\frac{\pzkuml[5]}{120}\He[4]{-\sqrt{\ssiz}\psnrh}
-\frac{\pzkuml[3]\pzkuml[4]}{24}\wrapBracks{\He[4]{-\sqrt{\ssiz}\psnrh}+\He[2]{-\sqrt{\ssiz}\psnrh}}
\right.\\
&\phantom{=}\,\,\,\,\phantom{+\oneby{\ssiz^{2}}}\left.
+\frac{\pzkuml[3]^3}{324}\wrapBracks{12\He[4]{-\sqrt{\ssiz}\psnrh} +
19\He[2]{-\sqrt{\ssiz}\psnrh}}\right]\\
&\phantom{=}\,\,+\ldots
\end{split}
\end{equation}
While this defines \psnrh implicitly, truncation gives polynomial
equations, whose roots can be found analytically or numerically.  Noting that
derivatives of Hermite polynomials can be easily computed, solving iteratively
for \psnrh via Newton's method should be simple. 

Truncating at two terms gives an equation which is quadratic in \psnrh,
yielding the (aesthetically unpleasing) solution:
\begin{equation}
\label{eqn:approx_hosnr_qfor}
\psnrh \approx \frac{3}{\pzkuml[3]} \pm \sqrt{\frac{9}{\pzkuml[3]^2} +
\frac{1}{\ssiz} - \frac{6\psnr}{\pzkuml[3]}}.
\end{equation}

As an example, for garden variety applications in asset management, setting
$\psnr=0.07\dayto{-\halff}$, $\pzkuml[3]=\ensuremath{-1}$,
$\ssiz=60\dayto{}$, we have $\psnrh\approx0.0719\dayto{-\halff}$.
If we consider a longer horizon, say $\ssiz=252\dayto{}$, one observes
$\psnrh\approx0.0698\dayto{-\halff}$.
Thus the difference between \psnrh and \psnr is modest at the
quarter year time scale, but negligible at the annual time scale. Note that
at the shorter time scale, $\sqrt{\ssiz}\psnr < 1$, resulting in a boost
to \psnrh due to negative skew, while at the longer time horizon, $\psnrh <
\psnr$ since $\sqrt{\ssiz}\psnr > 1$.


\section{Discussion}

It is not the purpose of this note to suggest that investors \emph{should}
optimize \psnrh. 
\emph{Prima facie}, the generalized Roy's criterion appears
inconsistent with the received wisdom that investors should maximize
expected utility, or corresponds somehow to decreasing risk
aversion\footnote{Perhaps Roy's criterion can be expressed in the classical
framework as a Heaviside utility function.}. 
Moreover, since Roy's criterion dichotomizes future returns, it shares some
of the hallmark failings of the Value at Risk measure, \viz that it does not
control for severe tail losses, may not promote diversification, \etc  \cite{delbaen2000}
Note, however, that Roy was decidely unenthusiastic about the prospect of 
maximizing expected utility, for pragmatic and philosophical reasons, writing,
``a man who seeks advice about his actions will not be grateful for the
suggestion that he maximise expected utility.''  \cite[p. 433]{roySafety1952}

While we do not have positive proof of investors who \emph{do} maximize Roy's
criterion, we can easily imagine there are some who might.
For example, at times a professional portfolio manager might 
try to maximize the probability of beating their benchmark over the next month,
fearing withdrawals from their fund\footnote{The title of this paper alludes
to this possible mismatch between goals of a fund investor and the fund manager:
in occupational safety, the ``Safety Third'' principle states that no party is as 
concerned with your personal wellbeing as you yourself are, with the implication that
overreliance on implicit workplace safeguards can be hazardous.}.
While investors cannot easily estimate, \emph{ex post}, what the 
\emph{ex ante} expected return of an investment should have been, they 
do exhibit a tendency to dichotomize their holdings as `winners' or `losers'.

Optimization of Roy's criterion provides an interesting mechanism by which
fully informed agents can agree on all moments of returns of an instrument,
yet rank the instrument differently based entirely on term. The short term
investor essentially sells (or leases, really) positive skew to the long term 
investor. It is not at all clear, however, that this differential preference
for skew drives the classical narrative of `investors' versus `speculators'; 
perhaps these two mythical groups can be separated by their appetite for kurtosis.

Finally, as a practical matter, it must be noted that maximization of Roy's criterion
is largely a quixotic pursuit. As illustrated in the sample calculation above,
the difference between \psnr and \psnrh tends to be small, much smaller in the
estimation error around \psnrh. Involving estimates of the higher order moments
of the returns distribution will only increase that estimation error.
\cite{lo2002,mertens2002comments,pav2015upsilon}

\nocite{haley2013smoothed}


\bibliographystyle{plainnat}
\bibliography{common,rauto}

\begin{thebibliography}{20}
\providecommand{\natexlab}[1]{#1}
\providecommand{\url}[1]{\texttt{#1}}
\expandafter\ifx\csname urlstyle\endcsname\relax
  \providecommand{\doi}[1]{doi: #1}\else
  \providecommand{\doi}{doi: \begingroup \urlstyle{rm}\Url}\fi

\bibitem[Abramowitz and Stegun(1964)]{abramowitz_stegun}
Milton Abramowitz and Irene~A. Stegun.
\newblock \emph{Handbook of Mathematical Functions with Formulas, Graphs, and
  Mathematical Tables}.
\newblock Dover, New York, ninth dover printing, tenth gpo printing edition,
  1964.
\newblock URL \url{http://people.math.sfu.ca/~cbm/aands/toc.htm}.

\bibitem[Chernozhukov et~al.(2007)Chernozhukov, Fern\'{a}ndez-Val, and
  Galichon]{edgeworthCF}
Victor Chernozhukov, Iv\'{a}n Fern\'{a}ndez-Val, and Alfred Galichon.
\newblock Rearranging {E}dgeworth-{C}ornish-{F}isher expansions.
\newblock Privately Published, 2007.
\newblock URL \url{http://arxiv.org/abs/0708.1627}.

\bibitem[Delbaen(2000)]{delbaen2000}
Freddy Delbaen.
\newblock Coherent risk measures.
\newblock 2000.
\newblock URL
  \url{https://people.math.ethz.ch/~delbaen/ftp/preprints/PISA007.pdf}.
\newblock Draft.

\bibitem[Denuit and Rey(2010)]{denuitprudence}
Michel Denuit and Béatrice Rey.
\newblock Prudence, temperance, edginess, and risk apportionment as decreasing
  sensitivity to detrimental changes.
\newblock 2010.
\newblock URL
  \url{http://sites.uclouvain.be/IAP-Stat-Phase-V-VI/PhaseVI/publications_2010/TR/TR10024.pdf}.

\bibitem[Eeckhoudt et~al.(2005)Eeckhoudt, Gollier, and
  Schlesinger]{eeckhoudt2005economic}
Louis Eeckhoudt, Christian Gollier, and Harris Schlesinger.
\newblock \emph{Economic and financial decisions under risk}.
\newblock Princeton University Press, 2005.
\newblock URL \url{http://idei.fr/doc/by/gollier/economic_financial.pdf}.

\bibitem[Haley et~al.(2013)Haley, Paarsch, and Whiteman]{haley2013smoothed}
M~Ryan Haley, Harry~J Paarsch, and Charles~H Whiteman.
\newblock Smoothed safety first and the holding of assets.
\newblock \emph{Quantitative Finance}, 13\penalty0 (2):\penalty0 167--176,
  2013.
\newblock URL \url{http://vinci.cs.uiowa.edu/~hjp/download/ssfrr.pdf}.

\bibitem[Hodges and Centre(1998)]{hodges1998generalization}
S.~Hodges and Financial Options~Research Centre.
\newblock \emph{A Generalization of the Sharpe Ratio and Its Applications to
  Valuation Bounds and Risk Measures}.
\newblock FORC preprint: Financial Options Research Centre. Financial Options
  Research Centre, Warwick Business School, University of Warwick, 1998.
\newblock URL
  \url{http://www2.warwick.ac.uk/fac/soc/wbs/subjects/finance/research/wpaperseries/1998/98-88.pdf}.

\bibitem[Jaschke(2001)]{Jaschke01}
Stefan~R. Jaschke.
\newblock The {C}ornish-{F}isher-expansion in the context of {D}elta - {G}amma
  - {N}ormal approximations.
\newblock Technical Report 2001,54, Humboldt University of Berlin,
  Interdisciplinary Research Project 373: Quantification and Simulation of
  Economic Processes, 2001.
\newblock URL \url{http://www.jaschke-net.de/papers/CoFi.pdf}.

\bibitem[Lee and Lin(1992)]{AS269}
Yoong-Sin Lee and Ting-Kwong Lin.
\newblock Algorithm {AS} 269: High order {C}ornish-{F}isher expansion.
\newblock \emph{Journal of the Royal Statistical Society. Series C (Applied
  Statistics)}, 41\penalty0 (1):\penalty0 pp. 233--240, 1992.
\newblock ISSN 00359254.
\newblock URL \url{http://www.jstor.org/stable/2347649}.

\bibitem[Lo(2002)]{lo2002}
Andrew~W. Lo.
\newblock {The Statistics of Sharpe Ratios}.
\newblock \emph{Financial Analysts Journal}, 58\penalty0 (4), July/August 2002.
\newblock URL \url{http://ssrn.com/paper=377260}.

\bibitem[Mertens(2002)]{mertens2002comments}
Elmar Mertens.
\newblock Comments on variance of the {IID} estimator in {Lo} (2002).
\newblock Technical report, Working Paper University of Basel,
  Wirtschaftswissenschaftliches Zentrum, Department of Finance, 2002.
\newblock URL
  \url{http://www.elmarmertens.com/research/discussion/soprano01.pdf}.

\bibitem[Pav(2015)]{pav2015upsilon}
Steven~E. Pav.
\newblock Inference on the {S}harpe ratio via the upsilon distribution.
\newblock Privately Published, 2015.
\newblock URL \url{http://arxiv.org/abs/1505.00829}.

\bibitem[Pratt(1964)]{pratt1964}
John~W. Pratt.
\newblock Risk aversion in the small and in the large.
\newblock \emph{Econometrica}, 32\penalty0 (1/2):\penalty0 pp. 122--136, 1964.
\newblock ISSN 00129682.
\newblock URL
  \url{http://www.aae.wisc.edu/aae706_content/References/Pratt-1964.pdf}.

\bibitem[Roy(1952)]{roySafety1952}
A.~D. Roy.
\newblock Safety first and the holding of assets.
\newblock \emph{Econometrica}, 20\penalty0 (3):\penalty0 pp. 431--449, 1952.
\newblock ISSN 00129682.
\newblock URL \url{http://www.jstor.org/stable/1907413}.

\bibitem[Sharpe(1965)]{Sharpe:1966}
William~F. Sharpe.
\newblock Mutual fund performance.
\newblock \emph{Journal of Business}, 39:\penalty0 119, 1965.
\newblock URL \url{http://ideas.repec.org/a/ucp/jnlbus/v39y1965p119.html}.

\bibitem[Smetters and Zhang(2013)]{NBERw19500}
Kent Smetters and Xingtan Zhang.
\newblock A sharper ratio: A general measure for correctly ranking non-normal
  investment risks.
\newblock Working Paper 19500, National Bureau of Economic Research, October
  2013.
\newblock URL \url{http://www.nber.org/papers/w19500}.

\bibitem[Stigler(1980)]{stiglersLaw}
Stephen~M. Stigler.
\newblock {S}tigler's law of eponymy.
\newblock \emph{Transactions of the New York Academy of Sciences}, 39\penalty0
  (1 Series II):\penalty0 147--157, 1980.
\newblock ISSN 2164-0947.
\newblock \doi{10.1111/j.2164-0947.1980.tb02775.x}.
\newblock URL \url{http://dx.doi.org/10.1111/j.2164-0947.1980.tb02775.x}.

\bibitem[Sullivan(2011)]{royTheMan}
Edward~J. Sullivan.
\newblock {A}.{D}. {R}oy: The forgotten father of portfolio theory.
\newblock \emph{Research in the History of Economic Thought and Methodology},
  29:\penalty0 pp. 73--82, 2011.
\newblock \doi{10.1108/S0743-4154(2011)000029A008}.
\newblock URL \url{http://www.emeraldinsight.com/books.htm?chapterid=1943472}.

\bibitem[Ulyanov(2014)]{CFetc}
Vladimir~V. Ulyanov.
\newblock {C}ornish – {F}isher expansions.
\newblock In Miodrag Lovric, editor, \emph{International Encyclopedia of
  Statistical Science}, pages 312--315. Springer Berlin Heidelberg, 2014.
\newblock ISBN 978-3-642-04897-5.
\newblock \doi{10.1007/978-3-642-04898-2_193}.
\newblock URL \url{http://dx.doi.org/10.1007/978-3-642-04898-2_193}.

\bibitem[Zakamouline and Koekebakker(2008)]{zakamulin2008portfolio}
Valeri Zakamouline and Steen Koekebakker.
\newblock {Portfolio Performance Evaluation with Generalized Sharpe Ratios:
  Beyond the Mean and Variance}.
\newblock \emph{SSRN eLibrary}, 2008.
\newblock \doi{10.2139/ssrn.1028715}.
\newblock URL \url{http://ssrn.com/paper=1028715}.

\end{thebibliography}

\appendix

\section{A counterexample}

Let \reti have mean and variance $\pmu>0$ and \psigsq, respectively. Let
\retj have the same distribution as \reti, except with probability $p > 0$
has an additional `bonus' return of a constant $B>0$. Clearly \retj
(first-order) stochastically dominates \reti. The mean of \retj is equal to
$\pmu + pB$.  The uncentered second moment of \retj is equal to $\psigsq +
\pmu^2 + pB^2$. The \txtSR of \retj is thus equal to
$$
\frac{\pmu + pB}{\sqrt{\psigsq -2\pmu pB -p^2B^2 + pB^2}}.
$$

Then if, for example, $\pmu=0.001, \psig=0.01, p=\ensuremath{10^{-4}},$ and
$B=0.25$, the \txtSR of \reti is $0.1$, while the \txtSR of
\retj is $0.0995$.

In fact, we can construct a sufficient condition for the \txtSR to be reversed
in this case. Since \pmu, $p$ and $B$ are assumed positive,
\begin{equation}
\begin{split}
&\frac{\pmu + pB}{\sqrt{\psigsq -2\pmu pB -p^2B^2 + pB^2}} \le \frac{\pmu}{\psig},\\
\Leftrightarrow \,& \frac{\wrapParens{\pmu + pB}^2}{\psigsq -2\pmu pB -p^2B^2 + pB^2} \le \frac{\pmu^2}{\psigsq},\\
\Leftrightarrow \,& \psigsq \wrapParens{\pmu + pB}^2 \le \pmu^2 \wrapParens{\psigsq -2\pmu pB -p^2B^2 + pB^2},\\
\Leftrightarrow \,& \psigsq \wrapParens{2pB\pmu + p^2B^2} \le \pmu^2 \wrapParens{-2\pmu pB -p^2B^2 + pB^2},\\
\Leftrightarrow \,& \psigsq \wrapParens{2\pmu + pB} \le \pmu^2 \wrapParens{-2\pmu -pB + B},\\
\Leftrightarrow \,& \wrapParens{\psigsq + \pmu^2}\wrapParens{2\pmu + pB} \le B \pmu^2,\\
\Leftrightarrow \,& \frac{2\pmu + pB}{B} \le \frac{\pmu^2}{\psigsq + \pmu^2},\\
\Leftrightarrow \,& p \le \frac{\pmu^2}{\psigsq + \pmu^2} - \frac{2\mu}{B}.
\end{split}
\end{equation}
In order for this last inequality to admit a solution with positive $p$, one
must have 
\begin{equation*}
\frac{B}{2} \ge \pmu + \frac{\psigsq}{\pmu}.
\end{equation*}
For the example above, this `minimum' value of $B$ is $0.202$, while the
maximum acceptable value for $p$ is $0.0019$.


\end{document}